\documentclass[journal=jacsat,manuscript=article]{achemso}
\usepackage[T1]{fontenc} 
\usepackage{epsfig}
\usepackage{float}
\usepackage{amsmath}
\usepackage{subfig}

\author{Ra\'ul Fuentes-Azcatl}
\affiliation{Instituto de F\'isica “Luis Rivera Terrazas”, Benem\'erita Universidad Aut\'noma de  Puebla, Apdo. Postal J-48, Puebla, 72570, Mexico}
\email{rfuentes@correo.xoc.uam.mx}

\title {Dielectric and Structural Study of Water with Various NaCl Concentrations via Molecular Dynamics
}

\begin{document}

\begin{abstract}

In this work, we investigate the structural and dielectric behaviour of water in sodium chloride (NaCl) solutions, ranging from low concentrations to levels approaching the solubility limit. Specifically, we examine concentrations from [NaCl] = 1m to [NaCl] = 6m, where 6.1m represents the solubility threshold of NaCl in H$_2$O.

For the water model, we employ the flexible TIP4P/$\epsilon_{Flex}$ model, which offers an enhanced reproduction of various properties compared to other flexible models and non-polarisable rigid models. This includes improvements in the prediction of the infrared spectrum. For the ions, we utilise two distinct models: NaCl/Madrid and NaCl/$\epsilon$, allowing for a comparative analysis of their impact on the system’s behaviour.

\end{abstract}

\section{Introduction}
Electrolyte solutions play a crucial role in facilitating many essential chemical and biological processes\cite{GRIRA}. Of particular interest are sodium chloride (NaCl) solutions, as these form the basis of environments like physiological serum and seawater, with some additional ions at lower concentrations. The behaviour of NaCl in solution is key to determining many physicochemical properties of these liquids. In recent decades, molecular simulation has become a valuable tool for investigating the properties of such systems. When a force field is properly validated, it can provide insights into conditions that may not be easily accessible experimentally. Additionally, numerical simulations offer detailed information on microscopic properties that are otherwise challenging to measure directly.

This is particularly significant in the context of electrolyte solutions, where experimental data generally describe the system as a whole but are often decomposed into single-ion contributions, which are sometimes reported as "experimental" values. A key advantage of molecular simulation is its ability to evaluate not only the global properties of the solution but also individual ionic contributions\cite{1}, though this may not always be unambiguous\cite{2}. Nevertheless, the quality of any simulation is inherently tied to the accuracy of the underlying force field. 

Simulations of electrolyte solutions date back several decades, and with recent advancements in computational power, the field has gained renewed interest. Early research was based on implicit solvent models, but the development of reliable water models and force fields that explicitly treat the solvent has led to more realistic representations of ionic behaviour. A significant body of research has focused on optimising the parameters for water-water, ion-ion, and ion-water interactions. One crucial aspect is the choice of water model, which is integral to the accuracy of the simulation\cite{Joung2009}.

Although non-polarisable force fields have traditionally been used for these simulations due to their computational efficiency, polarisation is an important factor to consider, especially in systems containing charged ions\cite{SmithD}. Surprisingly, despite expectations, non-polarisable models often perform comparably to polarisable ones in terms of overall accuracy. This may be due to the fact that early polarisable models were not fine-tuned, but with the development of highly accurate polarisable water models, the field is evolving, and we may see a shift towards their more frequent use in the simulation of ionic solutions.

Currently, the non-polarisable SPC/E model is widely used for simulating electrolyte solutions, offering a reasonable representation of water properties, though it does have certain limitations. For instance, SPC/E underpredicts the temperature of maximum density at ambient pressure, which hinders its utility in investigating how ionic concentration affects the solution's density. The crucial step in constructing simulations is creating a force field that accurately describes ion-water interactions. Typically, the model's parameters are fitted against experimental data on density and structure at specific temperatures and pressures, allowing for comparisons between simulation results and experimental findings for thermodynamic and dynamic properties\cite{Joung2009,Gee2011,Fyta,Jensen2006}.

A common strategy in studying electrolyte solutions involves combining one model for water with another for salt, both calibrated against the properties of their respective pure systems. Solubility is a key property used to validate the accuracy of these models. When NaCl dissolves in water, it dissociates into sodium (Na\textsuperscript{+}) and chloride (Cl\textsuperscript{-}) ions, each surrounded by water molecules due to the polar nature of the solvent. At certain concentrations, the system undergoes phase separation into salt-rich and salt-poor phases, and the solubility can be computed by studying this coexistence.

Several methods exist to calculate solubility, one of which involves estimating the chemical potentials of both the solid and the saturated solution. For the solid phase, the absolute free energy can be determined using techniques such as the Ferrario et al.\cite{Ferrario} method. Sanz and Vega, for example, applied this approach to calculate the solubility of NaCl and potassium fluoride (KF) in water\cite{Sanz}, with subsequent studies expanding this methodology to other salts. Another approach uses large crystals in contact with an almost saturated ion solution to reach equilibrium, though this method requires very long simulations and large systems to minimise finite-size effects.\cite{manzanilla}

One of the challenges in studying salt-water mixtures using separate water and salt models is that when water surrounds the ions, it influences the interactions between the salt and water molecules, affecting both water-water and salt-salt interactions.

In this study, we employ the flexible TIP4P/$\epsilon_{Flex}$\cite{tip4pef} water model, which provides a significantly improved reproduction of various physical properties compared to both other flexible models and non-polarisable rigid models. One of the key advantages of TIP4P/$\epsilon_{Flex}$ is its ability to more accurately predict the infrared spectrum, a crucial aspect in understanding molecular dynamics and interactions within the system. 

For the ionic components, we utilise two distinct force field models: NaCl/Madrid \cite{Benavides} and NaCl/$\epsilon$\cite{nacle}. The inclusion of these two models allows for a comprehensive comparative analysis, enabling us to assess their respective influences on the structural and dielectric behaviour of the system. The NaCl/$\epsilon$\cite{nacle} that is parameterised to reproduce the properties of both the pure salt and the water-salt mixture. Is tested the model’s performance against experimental data for both the pure salt and the solution, employing water models such as SPC/$\epsilon$\cite{spce} and TIP4P/$\epsilon$\cite{tip4pe}, which are designed to give an accurate dielectric constant for water. This dual-model approach is particularly valuable for investigating how variations in force field parameterisations can affect the simulation outcomes, especially in terms of the interaction between ions and the solvent. Through this, we aim to provide deeper insights into the accuracy and limitations of current models for simulating electrolyte solutions, particularly in the context of high ionic concentrations.

By combining the advancements in water models with a detailed exploration of salt-water interactions, this work aims to refine our understanding of the behaviour of NaCl solutions across a range of concentrations, from dilute solutions to near the solubility limit. This contributes to broader research efforts in the fields of chemical and biological processes, where electrolyte solutions are fundamental.

This paper is organised as follows. In Section 2, we introduce the force fields employed in this study. Section 3 outlines the simulation details, including the parameters and methodologies applied. The results and analysis are presented in Section 4, while Section 6 concludes with a discussion of the findings and their implications.

\section{The Force Fields}

\subsection{The Water Model}
For the intermolecular potential, we employed a four-site model TIP4P/$\epsilon _{flex}$ \cite{tip4pef}, akin to the widely used TIP4P model structures. This model features a Lennard-Jones centre at the position of the oxygen atom, which captures the van der Waals interactions between water molecules. Additionally, the electrostatic interactions are represented by two positive charges located at the hydrogen atoms and a compensating negative charge situated at the so-called M-site, which lies off the oxygen atom.

\begin{figure}[h!]
	{\includegraphics[width=0.7\textwidth,angle=0]{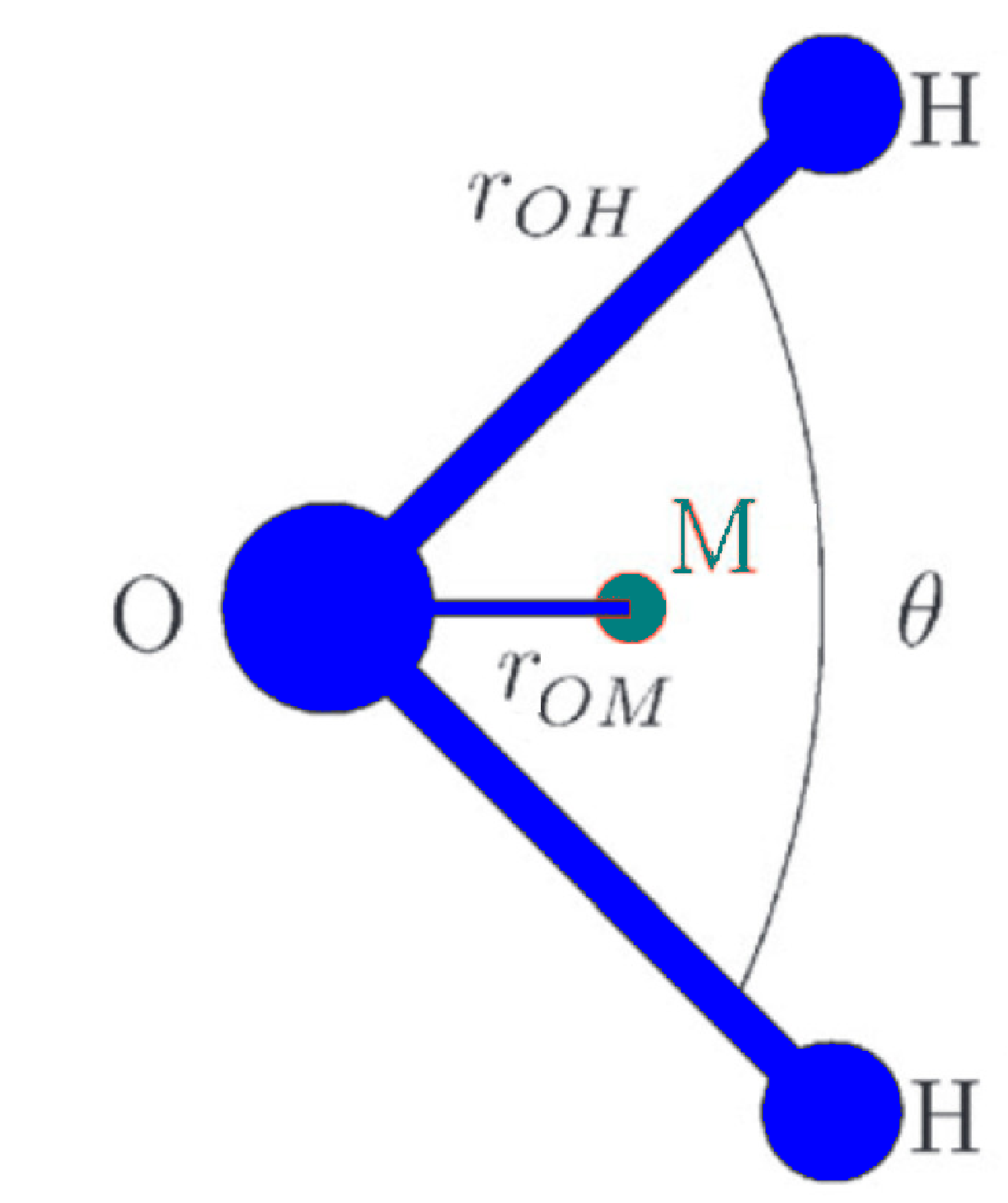}}
	
\caption{Schematic representation of the TIP4P family of water models. The distance between the oxygen atom and each hydrogen atom is denoted as \(r_{\text{OH}}\), and the bond angle between the oxygen and the two hydrogen atoms is given by \(\theta\). The hydrogen atoms bear a partial positive charge, while the negative charge is positioned at an off-centre location, designated as point M, at a distance \(r_{\text{OM}}\) from the oxygen atom. It is important to note that point M does not correspond to the actual position of any atom but is introduced to more accurately model the electrostatic properties of the water molecule.
}
		\label{mol-h2o}
	\end{figure}

In this configuration, water molecules consist of four interaction sites: two hydrogens, one oxygen, and the M-site. The M-site is positioned at a specific distance from the oxygen atom along the bisector of the hydrogen atoms, figure \ref{mol-h2o}. This setup accurately reflects the geometry of the water molecule and the distribution of its charge, making it highly suitable for simulating the behaviour of water in different phases and under various conditions.


The use of a flexible water model is key to capturing the inherent dynamic behaviour of water molecules, as opposed to rigid models that treat bond lengths and angles as fixed. In the flexible model, the intramolecular interactions are defined using harmonic potentials, which allow for realistic stretching of bonds and bending of angles between atoms. This flexibility is important because it accounts for the slight variations in bond lengths and angles caused by thermal motion and external influences, making the model more reflective of real-world molecular dynamics.

A flexible model of water is particularly useful in simulations where the dynamic behaviour of hydrogen bonds and the molecular flexibility of water play a critical role, such as in biological systems or phase transitions\cite{jorge}. Water’s ability to form and break hydrogen bonds dynamically is central to its unique properties, such as high surface tension, heat capacity, and solvent behaviour. Flexible models enable the simulation to capture these characteristics more accurately, as they allow for slight distortions in the O-H bonds and H-O-H angles in response to thermal fluctuations or interactions with other molecules.

Additionally, flexible models are crucial for systems where molecular vibrations and rotations significantly affect the outcome of the simulation, as they can adapt to changing conditions like temperature and pressure. This is particularly relevant when studying liquid water, where hydrogen bonding networks are constantly shifting, or when simulating the behaviour of water in extreme environments like supercooled water or high-pressure conditions. In these cases, flexibility allows the model to account for the vibrational and rotational degrees of freedom that rigid models might miss, resulting in more accurate predictions of thermodynamic properties, such as diffusion coefficients, dielectric constants, and phase behaviour\cite{tip4pef,fbae}.

The intramolecular interactions in the flexible model are defined
by harmonic potentials in bonds and angle,

\begin{equation}
\label{k}
U_k(r)=\frac{k_r}{2}(r-r_0 )^2 
\end{equation}
\noindent and in the H-O-H angle,
\begin{equation}
\label{theta}
U_{\theta}(\theta)=\frac{k_{\theta}}{2}(\theta-\theta_0)^2 ,
\end{equation}
\noindent where $r$ is the bond distance and $\theta$ is the bond angle. The subscript $0$ denotes their equilibrium  values, and $k_r$ and $k_{\theta}$ are the corresponding spring constants. 

For the intermolecular potential between two molecules the LJ and Coulomb interactions are used for non-polarizable models,
\begin{equation}
\label{ff}
u(r) = 4\epsilon_{\alpha \beta} 
\left[\left(\frac {\sigma_{\alpha \beta}}{r}\right)^{12}-
\left (\frac{\sigma_{\alpha \beta}}{r}\right)^6\right] +
\frac{1}{4\pi\varepsilon_0}\frac{q_{\alpha} q_{\beta}}{r}
\end{equation}

\noindent where $r$ is the distance between sites $\alpha$ and $\beta$, $q_\alpha$ is the electric charge of site $\alpha$, $\varepsilon_0$ is the permitivity of vacuum,  $\epsilon_{\alpha \beta}$ is the LJ energy scale and  $\sigma_{\alpha \beta}$ the repulsive diameter for an $\alpha-\beta$ pair. The cross interactions between unlike atoms are obtained using the Lorentz-Berthelot mixing rules,

\begin{equation}
\label{lb}
\sigma_{\alpha\beta} = \left(\frac{\sigma_{\alpha\alpha} +
	\sigma_{\beta\beta} }{2}\right);\hspace{1.0cm} \epsilon_{\alpha\beta} =
\left(\epsilon_{\alpha\alpha} \epsilon_{\beta\beta}\right)^{1/2}
\end{equation}
The geometry and parameters of the Force Fields for the TIP4P/$\varepsilon_{Flex}$ are given in the table \ref{tip4pe}.


\begin{table}[h]
\caption{Force field parameters of TIP4P/$\varepsilon_{Flex}$ water model. The negative charge in site $M$ is $q_M=-(2 q_H)$.  }
\label{tip4pe}
\begin{center}
\begin{tabular}[h]{|cccccccccc|}
\hline
\hline Model & k$_b$& $r_{OH_{eq}}$ &k$_a$& $\Theta_{eq}$  & $q_H$  & $q_M$& $d{^{rel}}_{OM}$   & $\sigma$ & $(\epsilon/k_B)$\\

& kJ/ mol\AA² &  \AA & kJ/ mol rad²&  $^0$  & e  & e & \AA  & \AA & K\\

\hline

TIP4P/$\varepsilon_{Flex}$& 1570&0.9300&212& 111.50 &0.51& -1.02& 0.0830 &  3.1734 & 95.5\\

\hline
\end{tabular}
\end {center}
\end{table}

\subsection{NaCl models}

One of the principal limitations of molecular dynamics simulations lies in the fact that force fields do not always accurately capture the diversity and complexity of experimental data. This can result in deviations from real-world observations and, in some cases, lead to incorrect or misleading conclusions about the system under study. For instance, certain physical and thermodynamic properties, such as phase behaviour, structural features, and dynamic responses, may be inadequately represented.

The NaCl/$\epsilon$\cite{nacle} model has demonstrated a notable improvement in reproducing a wide range of properties in both the liquid and solid phases of sodium chloride. This model enhances the accuracy of predictions by better reflecting the behaviour of NaCl in these states. Additionally, when this model is applied to NaCl mixed with water, it significantly improves the reproduction of key thermodynamic properties of the mixture across various concentrations. This makes it particularly useful for simulations where salt-water interactions are critical, such as in biological systems or industrial processes.

Our research group has made substantial progress in developing more refined and reliable force fields. Through careful adjustments and testing, we have managed to adapt these improved force fields to other models with minimal modifications\cite{ILe,CO2e,KBre} of the new parameters, which is critical for broadening their application across different systems.

Specifically, the parameters for the NaCl/$\epsilon$ model were meticulously fitted to reproduce the experimental density of crystalline NaCl at standard conditions, a temperature of 298 K and a pressure of 1 bar, yielding a value of 2.16 g cm$^{-3}$. This level of accuracy is maintained over a range of temperatures, showcasing the model’s reliability and flexibility. Furthermore, the model effectively reproduces surface tension, a challenging property to model accurately, which further underscores its strength in simulating interfacial phenomena.

In addition to the NaCl/$\epsilon$ model, we employed another force field proposed by Benavides et al.\cite{Benavides}, which is an adaptation of the $\epsilon$ model specifically tailored to the TIP4P/05 water model. This adaptation is particularly significant as it calculates solubility based on the chemical potential, providing a more precise representation of the thermodynamics of dissolution. Table \ref{nacle} presents the force field parameters for both the NaCl/$\epsilon$ model and the NaCl/Madrid model.It is worth noting that none of the ion models were parametrised using the flexible water model. Therefore, an analysis of their transferability is also conducted to evaluate how well the parameters, originally developed for rigid models, perform when applied to the flexible TIP4P/$\epsilon_{Flex}$ framework.

\begin{table}							
	\caption{ Force Field Parameter of NaCl/$\epsilon$ and NaCl/Madrid }
	\label{nacle}       							
	\begin{tabular}{|cccc|}       							
		\hline							
		model	&	q/e	&	$\sigma$/Å	&	($\epsilon$/k$_B$)/K	\\
		\hline
		NaCl/$\epsilon$ :& & & \\
									
		Na	&	0.885	&	2.52	&	17.44	\\
		Cl	&	-0.885	&	3.85	&	192.45	\\
		\hline      
		\hline  							
		NaCl/Madrid :& & & \\
		Na	&	0.85	&	2.2174	&	177.086	\\
		Cl	&	-0.85	&	4.8490	&	9.2518\\
		\hline       							
	\end{tabular}       							
\end{table}

\section{Simulation Details}

Molecular dynamics (MD) simulations were carried out using GROMACS (version 2022)\cite{Pall15}. The equations of motion were integrated using the leapfrog algorithm~\cite{Allen89, Pall15}, with a time step of 1 femtosecond (fs). Each simulation ran for 250 nanoseconds (ns) in each molalities, and the positions and velocities of the particles were recorded every 1000 steps to capture the system’s evolution over time.

To account for long-range electrostatic interactions, Ewald summations were employed. The real-space part of the Coulombic potential was truncated at 10 Å. The Fourier component of the Ewald sums was calculated using the smooth particle mesh Ewald (SPME) method\cite{Essman95}, with a grid spacing of 1.2 Å. A fourth-degree polynomial interpolation was applied to ensure precise and smooth evaluation of the electrostatic contributions in reciprocal space.

The simulation box was maintained as a cubic structure throughout the entire simulation, providing consistency in the boundary conditions. Since the model is flexible, no constraints are applied to the bond angles or bond lengths. 

The temperature was controlled using a Nos\'e-Hoover thermostat\cite{Tuckerman01}. The thermostat relaxation time constant, $\tau_T$, was set to 0.4 picoseconds (ps), allowing the system to equilibrate to the desired temperature efficiently while maintaining accurate thermal fluctuations. The pressure was regulated using the Parrinello-Rahman barostat, which is widely used for constant pressure (NPT) simulations. The barostat relaxation time constant, $\tau_P$, was set to 0.4 ps, ensuring stable pressure control throughout the simulation.

For sodium chloride (NaCl/$\epsilon$, NaCl/Madrid) in water, the simulations were conducted using a system consisting of 864 molecules. Under various conditions, the molal concentration exhibits notable variations, as illustrated in Table \ref{molal}. The simulations were carried out across a range of molalities, while maintaining a constant temperature of 298 K and a pressure of 1 bar, representative of standard conditions for studying aqueous solutions at ambient temperature. The utilisation of the NPT ensemble allows for fluctuations in both volume and pressure, ensuring that thermodynamic properties are accurately represented, particularly in systems involving liquids and dissolved solutes. 

In the case of pure water, it has been demonstrated that a system size of 250 molecules and a cutoff radius of 0.7 nm is sufficient to reproduce key properties such as density and dielectric constant with high precision \cite{diel}.

The molality concentration of the solution was determined from the total number of ions in solution,N$_{ions}$,the number of water molecules, N$_{H2O}$, and the molar mass of water, M$_{H2O}$. Molality, which is defined as the number of moles of solute per kilogram of solvent, is an important measure in studying electrolyte solutions like NaCl, as it provides a direct relationship between the number of dissolved particles and the solvent mass. The molality was calculated using the following equation:

For sodium chloride (NaCl/$\varepsilon$) in water, the simulations were conducted using 864 molecules in the isothermal-isobaric (NPT) ensemble, in the liquid phase at various molalities under ambient conditions. The molality concentration is derived from the total number of ions in solution, $N_{\text{ions}}$, the number of water molecules, $N_{\text{H}_2\text{O}}$, and the molar mass of water, $M_{\text{H}_2\text{O}}$, as follows:

\begin{equation}
	\label{mol}
	\left[\text{NaCl}\right] = \frac{N_{\text{ions}} \times 10^{3}}{2N_{\text{H}_2\text{O}} M_{\text{H}_2\text{O}}} \; .
\end{equation}

In this equation, the division by 2 accounts for a pair of ions (Na$^{+}$ and Cl$^{-}$), and the molar mass of water is taken as $M_{\text{H}_2\text{O}} = 18$ g mol$^{-1}$. Table \ref{molal} provides the molality values for each point in the simulation, calculated using this formula.

\begin{table}[h]
	\caption{Composition of NaCl solutions used in the simulations at 298.15 K and 1 bar.}
	\label{molal}
	\begin{center}
		\begin{tabular}{|ccc|}
			\hline\hline
			Molality (m) & $N_{\text{H}_2\text{O}}$ & $N_{\text{ions}}$ \\
			\hline
			0.99 & 832 & 32 \\
			1.99 & 806 & 58 \\
			3.07 & 778 & 86 \\
			4.05 & 754 & 110 \\
			5.00 & 732 & 132 \\
			6.02 & 710 & 154 \\
			\hline
		\end{tabular}
	\end{center}
\end{table}

This table lists the number of water molecules, $N_{\text{H}_2\text{O}}$, and the number of sodium and chloride ions, $N_{\text{ions}}$, corresponding to each molality, thus illustrating the specific compositions used in the simulations.

\section{Results}

The thermodynamic and dynamic properties of NaCl (using both the $\epsilon$ and Madrid models) in solution with TIP4P/$\epsilon_{Flex}$, water were compared against experimental results at various molal concentrations, after 250 ns of simulation.

\subsection{Density, $\rho$.}

The density of the mixture of NaCl and water as a function of the salt molal concentration at 1 bar and 298 K for NaCl/$\epsilon$ in TIP4P/$\epsilon_{Flex}$, NaCl/Madrid in TIP4P/$\epsilon_{Flex}$, and the experimental results is presented in Figure \ref{}.

It is important to note that there is a minimal difference between the TIP4P/05 and TIP4P/$\epsilon$ models (which were used to parameterise the Madrid and $\epsilon$ force fields of NaCl). The latter, TIP4P/$\epsilon$, can better reproduce the dielectric constant and the isothermal compressibility, while the other experimental values are similarly reproduced by both models. Therefore, the key distinction lies in the dielectric constant, which is somewhat underestimated in the TIP4P/05 model.

Consequently, when employing the flexible model, a significant difference is observed in the calculation of density, particularly in the combination of the NaCl/Madrid model with TIP4P/$\epsilon_{Flex}$. This difference underscores the sensitivity of density calculations to the model's ability to accurately capture dielectric properties, which in turn influences the behaviour of ion-solvent interactions in the solution.


\begin{figure}[h!]
	{\includegraphics[width=0.7\textwidth,angle=-90]{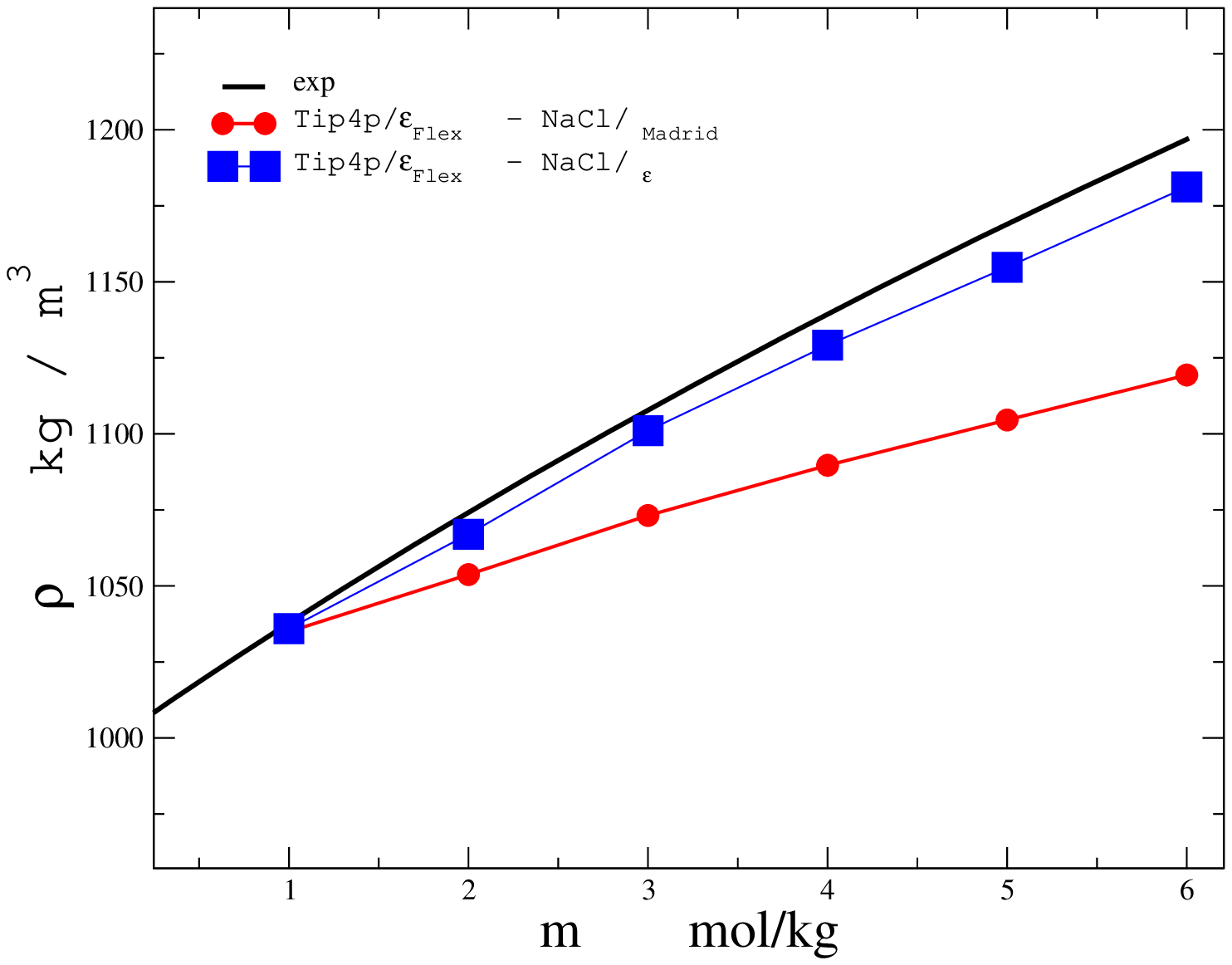}}
	
\caption{Density of the mixture versus molal concentration of the salt
at the temperature of 298 K and at 1 bar pressure. The black line is the
experimental data\cite{CRC}, for NaCl/$\epsilon$ in TIP4P/$\epsilon_{Flex}$water (blue squares) and NaCl/Madrid in TIP4P/$\epsilon_{Flex}$, water (red circles)}
		\label{Dens}
	\end{figure}
\newpage

\subsection{Isothermal Compressibility , $\kappa$.}

The TIP4P/$\epsilon_{Flex}$ model accurately reproduces the compressibility, denoted as $\kappa$, under conditions of 298 K and 1 bar of temperature and pressure, respectively. It is noteworthy that both force fields for NaCl result in higher values of compressibility at each molality, leading to an overestimation of this property. This indicates that the calculated isotherms are approximately one-fifth higher than the experimental values.

In solutions, the isotermic compressibility can significantly influence how solutes dissolve and behave within a solvent. An understanding of compressibility is essential, as it affects various phenomena such as solvation dynamics, reaction kinetics, and the stability of colloidal systems. Furthermore, accurate predictions of compressibility are crucial for the design of processes in chemical engineering, as they inform the behaviour of liquid mixtures under varying conditions of pressure and temperature.


\begin{figure}[h!]
	{\includegraphics[width=0.7\textwidth,angle=-90]{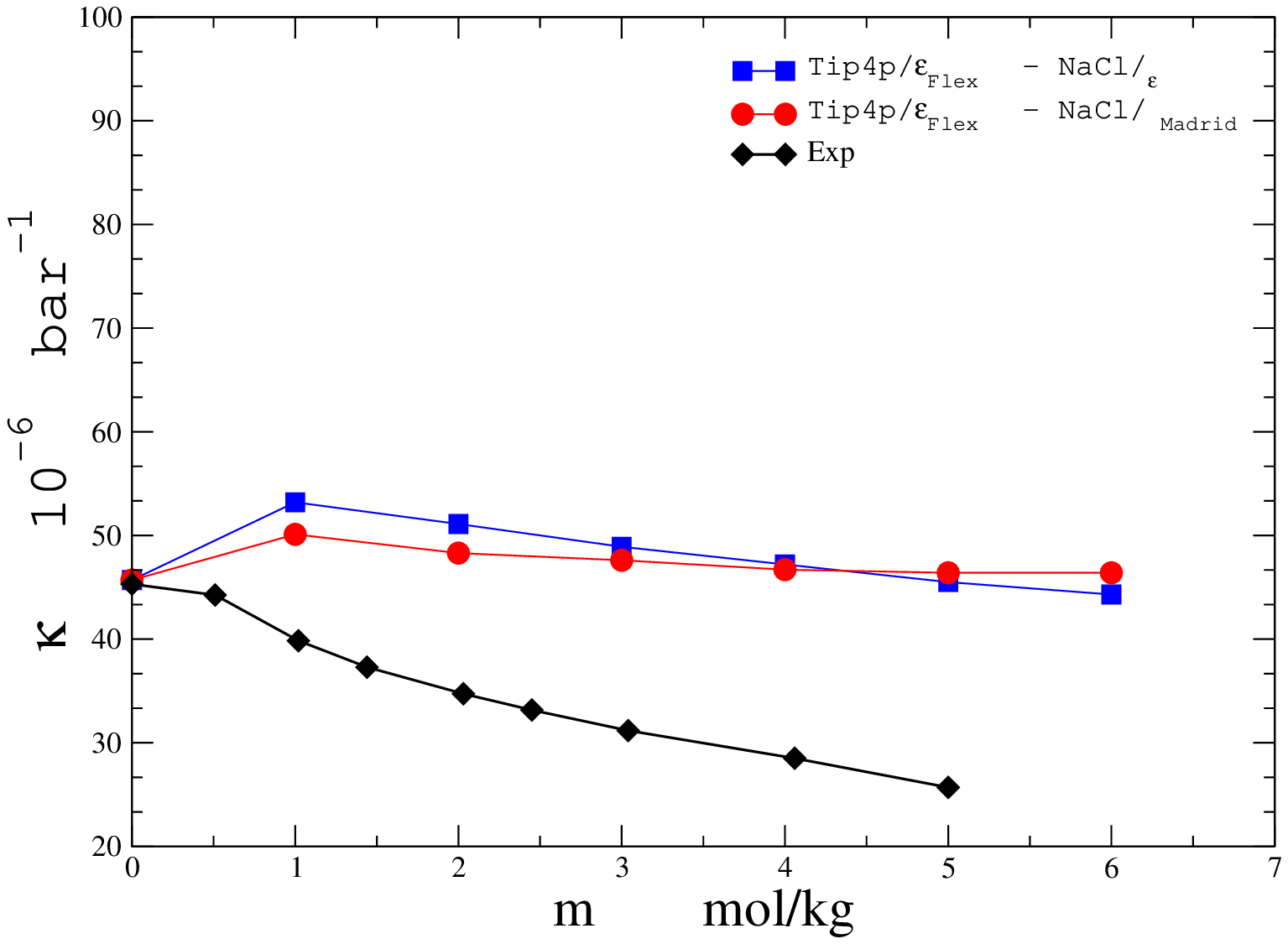}}
	
\caption{Isothermal Compressibility ($\kappa$) of the mixture versus molal concentration of the salt
at the temperature of 298 K and at 1 bar pressure. The black line is the
experimental data\cite{millero}, for NaCl/$\epsilon$ in TIP4P/$\epsilon_{Flex}$water (blue squares) and NaCl/Madrid in TIP4P/$\epsilon_{Flex}$, water (red circles)}
		\label{kapa}
	\end{figure}

\subsection{Larger Cluster and Free Ions}

In the study of electrolyte solutions and their behaviour, it is essential to consider the formation of larger clusters and the presence of free ions. Clusters refer to aggregates of particles or ions that arise due to intermolecular forces and electrostatic interactions. In the case of salt solutions, such as sodium chloride (NaCl), the ions dissociate in the aqueous medium, generating both free ions and clusters that can significantly influence the properties of the solution.

In this study, we analyse these clusters, ensuring that they do not reach a size that would disrupt the Gibbs free energy equilibrium, potentially affecting nuclear interactions. Monitoring the size of these clusters is crucial, as larger aggregates may lead to alterations in the thermodynamic properties of the solution. For instance, excessively large clusters can hinder the mobility of free ions, thereby impacting the conductivity and reactivity of the solution.

Furthermore, understanding the dynamics of cluster formation and dissociation can provide insights into the solvation processes and the overall stability of the solution. The balance between free ions and clusters is vital for accurately predicting the behaviour of electrolytes under various conditions, which is of paramount importance in fields such as physical chemistry, materials science, and biochemistry.

\begin{figure}[h!]
	{\includegraphics[width=0.7\textwidth,angle=-90]{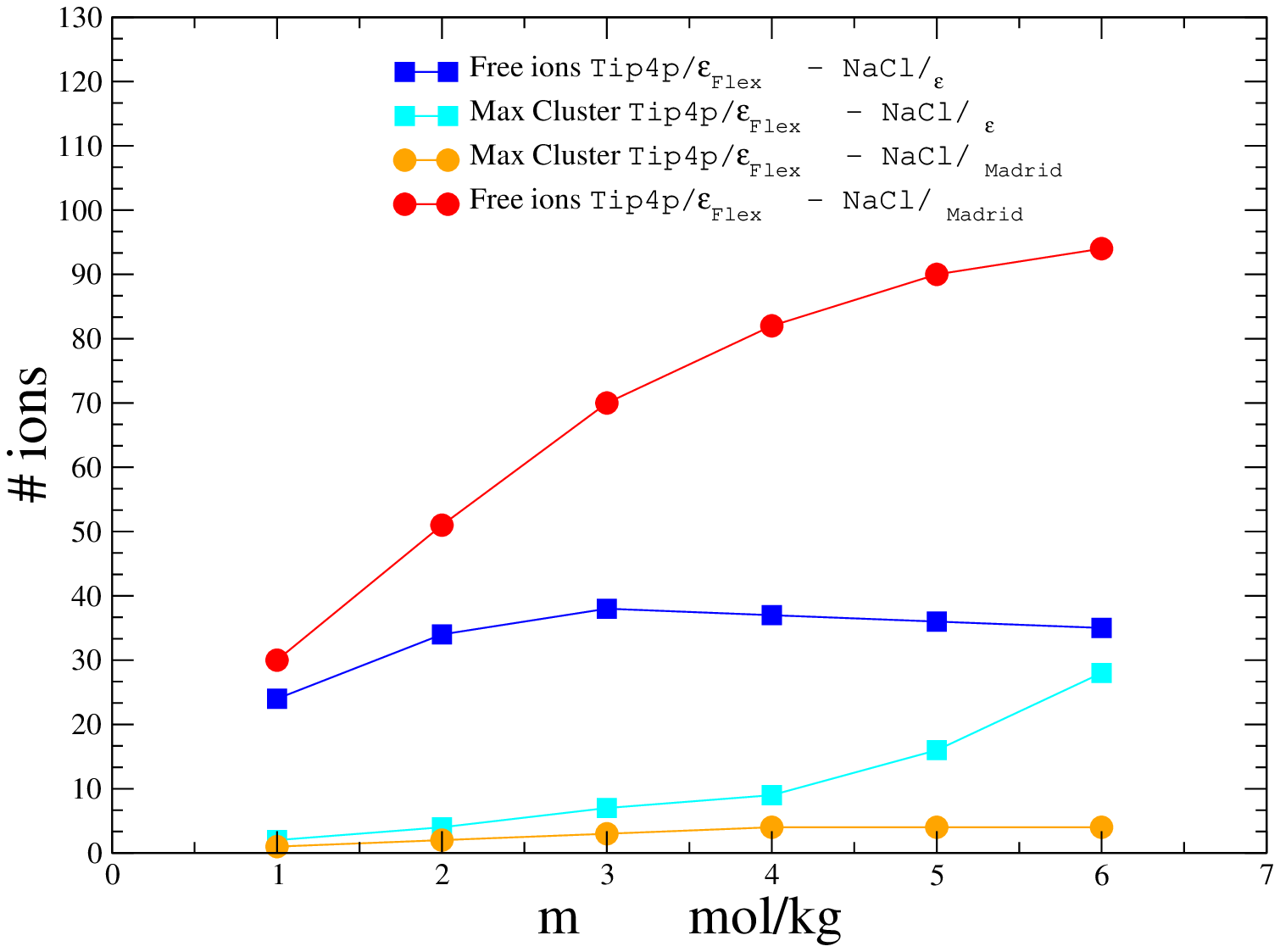}}
	
\caption{Free ions and larger cluster of the mixture versus molal concentration of the salt
at the temperature of 298 K and at 1 bar pressure. Free ions for NaCl/$\epsilon$ in TIP4P/$\epsilon_{Flex}$ water (blue squares) and NaCl/Madrid in TIP4P/$\epsilon_{Flex}$, water (red circles). The Larger cluster for NaCl/$\epsilon$ in TIP4P/$\epsilon_{Flex}$ water (cian squares) and NaCl/Madrid in TIP4P/$\epsilon_{Flex}$, water (orange circles)
}
		\label{cluster}
	\end{figure}

In Figure \ref{cluster}, we present the calculations for the free ions and the largest cluster formed on average throughout the simulation, which spans 250 ns in each case. We observe that the calculations carried out using the NaCl/Madrid force field in TIP4P/$\epsilon_{Flex}$ never result in the formation of a cluster larger than four ions, with free ions being present in greater proportion.

In contrast, the results obtained with NaCl/$\epsilon$ in TIP4P/$\epsilon_{Flex}$ indicate the formation of larger clusters, particularly from a molar concentration of [NaCl] = 3m, where the largest cluster begins to grow. Additionally, free ions tend to stabilise on average from a concentration of [NaCl] = 2m, suggesting the formation of small clusters throughout the simulation.

This behaviour underscores the differing aggregation dynamics between the two force fields, with NaCl/$\epsilon$ facilitating the formation of more substantial clusters at higher concentrations, while NaCl/Madrid predominantly maintains a higher ratio of free ions.


\newpage

\subsection{Fluctuation of the Dipole Moment, <M²>-<M>².}

The dipole moment of a molecule is a crucial parameter that quantifies the separation of positive and negative charges within the molecule, thereby reflecting its polarity. In molecular simulations, the fluctuation of the dipole moment is an important property to consider, as it provides insights into the molecular interactions and dynamics of the system under study.

\begin{figure}[h!]
	{\includegraphics[width=0.7\textwidth,angle=-90]{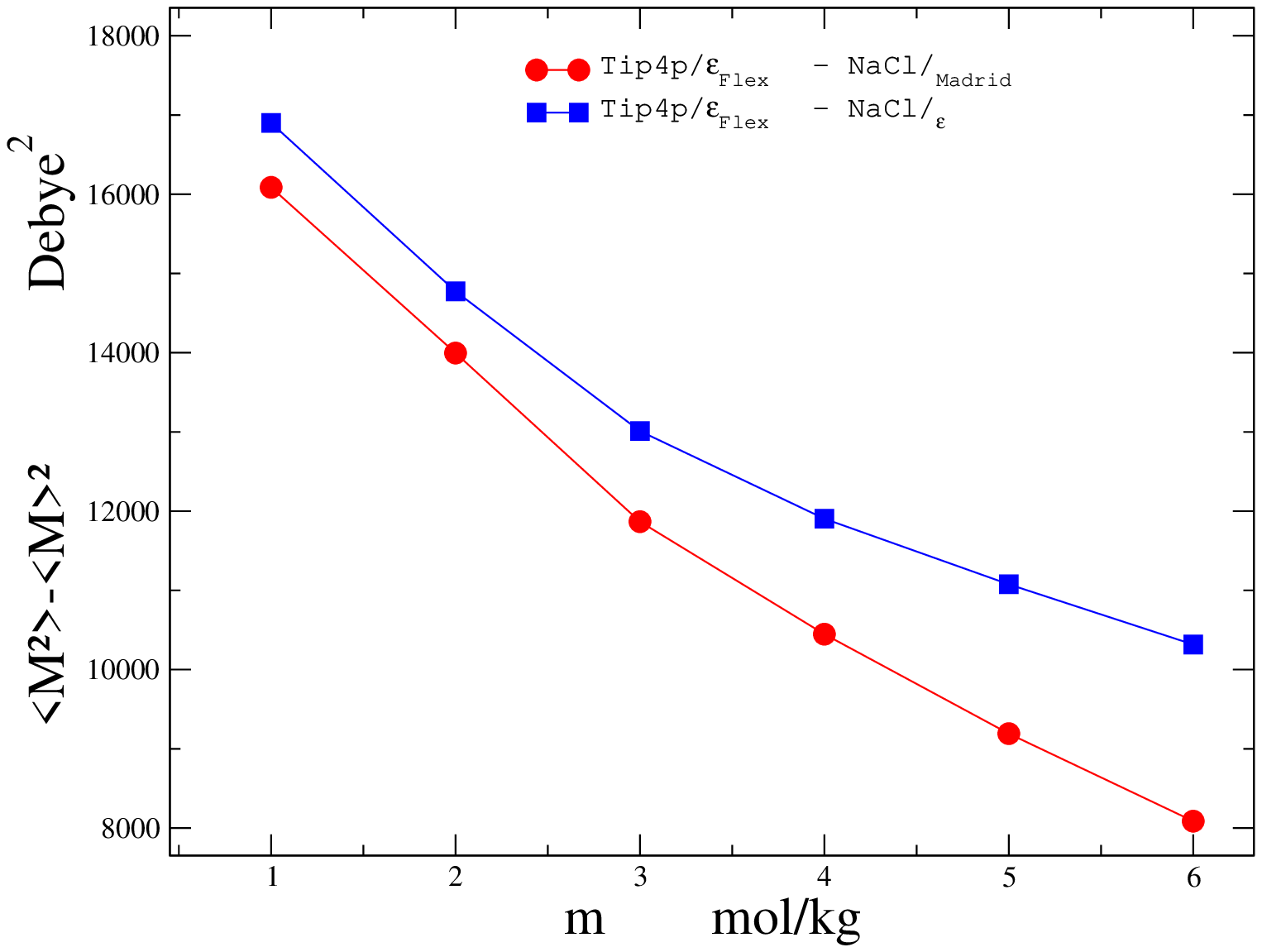}}
	
\caption{Fluctuation of the Dipole Moment of the mixture versus molal concentration of the salt
at the temperature of 298 K and at 1 bar pressure. The black line is the
experimental data, for NaCl/$\epsilon$ in TIP4P/$\epsilon_{Flex}$water (blue squares) and NaCl/Madrid in TIP4P/$\epsilon_{Flex}$, water (red circles)}
		\label{DM}
	\end{figure}

Understanding dipole moment fluctuations is essential for elucidating the dynamics of polar systems, as these fluctuations can affect properties such as dielectric constant, solvation energies, and reaction rates. Additionally, in ionic solutions, dipole fluctuations can influence ion pairing and solvation shell formation, thereby affecting the overall thermodynamic properties of the solution, see figure\ref{DM}.


\( M \) represents the total dipole moment \( \mu \) in the system. To quantify the extent of the fluctuations in this dipole moment, we can calculate the mean square fluctuation of the dipole moment, denoted as \( \langle (\Delta M)^2 \rangle \). This fluctuation can be expressed mathematically as:

\begin{equation}
\langle (\Delta M)^2 \rangle = \langle M^2 \rangle - \langle M \rangle^2
\end{equation}

In this equation, \( \langle M \rangle \) denotes the average dipole moment over the duration of the simulation. The term \( \langle M^2 \rangle \) represents the mean square of the dipole moment, while \( \langle M \rangle^2 \) is the square of the average dipole moment. 

By calculating \( \langle (\Delta M)^2 \rangle \), we can gain insights into the dynamics of the system, particularly how the dipole moment evolves over time in response to thermal fluctuations and intermolecular interactions. This analysis is essential for understanding the behaviour of polar systems and their dielectric properties.

\subsection{Dielectric constant, $\epsilon$.}
\begin{figure}[h!]
	{\includegraphics[width=0.7\textwidth,angle=-90]{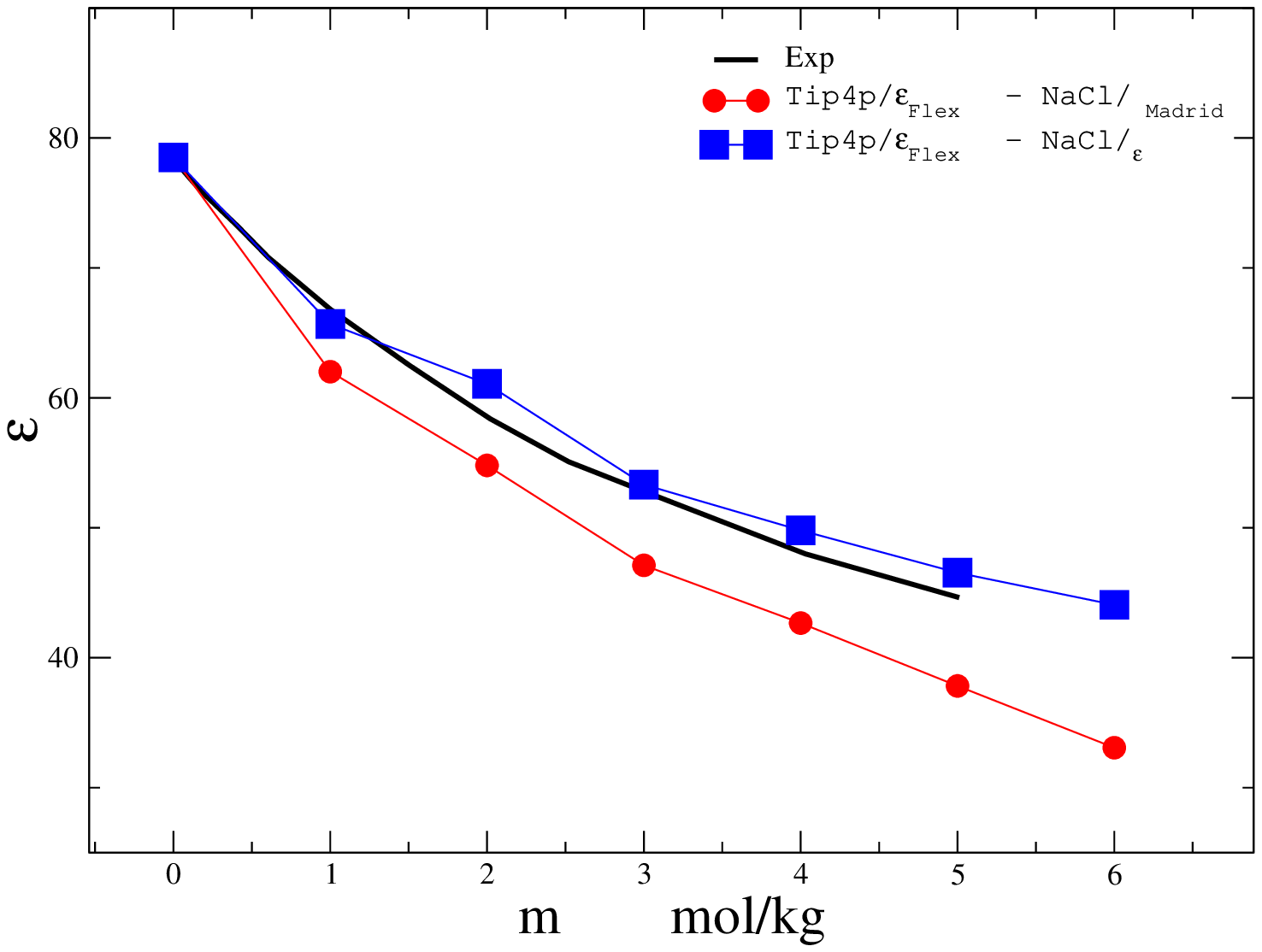}}
	
\caption{Dielectric constant of the mixture versus molal concentration of the salt
at the temperature of 298 K and at 1 bar pressure. The black line is the
experimental data\cite{CRC}, for NaCl/$\epsilon$ in TIP4P/$\epsilon_{Flex}$water (blue squares) and NaCl/Madrid in TIP4P/$\epsilon_{Flex}$, water (red circles)}
		\label{CteDiel}
	\end{figure}

 Figure \ref{CteDiel} presents the dielectric constant as a function of salt molal concentration at 298 K and 1 bar, for NaCl/$\epsilon$ in TIP4P/$\epsilon_{Flex}$water (blue squares) and NaCl/Madrid in TIP4P/$\epsilon_{Flex}$, water (red circles). These results are compared with experimental data (solid black line).
A significant deviation is observed in the NaCl/Madrid in TIP4P/$\epsilon_{Flex}$ , model at higher salt concentrations, which can be attributed primarily to the lower energetic interaction between the ions and water molecules in this model. This suggests that the Madrid model underestimates the strength of ion-water interactions, leading to a noticeable discrepancy with experimental values, particularly at elevated salt concentrations.
In contrast, the NaCl/$\epsilon$ model in TIP4P/$\epsilon_{Flex}$, water shows a much closer agreement with experimental data. At a concentration of 3 mol/kg, the error between the simulation and experimental results is reduced to 0.5\%, highlighting the improved accuracy of the NaCl/$\epsilon$ model in capturing the behaviour of the solution under these conditions.
The overall performance of the NaCl/$\epsilon$ model demonstrates its capability to replicate the dielectric properties of sodium chloride solutions with greater fidelity at elevated molalities. This improved performance is likely due to the more accurate representation of ion-water interactions in the $\epsilon$ model, which becomes increasingly important as ion concentration rises and intermolecular forces play a more significant role in determining the dielectric properties of the solution.


The static dielectric constant of a liquid can be determined from molecular dynamics (MD) trajectories using various methods. A widely adopted approach for calculating the dielectric constant of a bulk liquid involves analysing fluctuations in the total dipole moment, as described by Neumann~\cite{neumann}. This methodology utilises Equation \ref{diel} for the total dipole moment:

\begin{equation}
\label{diel}
\epsilon = 1 + \frac{1}{3k_B T V \epsilon_0} \left( \langle M^2 \rangle - \langle M \rangle^2 \right),
\end{equation}

where \( k_B \) is the Boltzmann constant, \( T \) is the absolute temperature, \( \epsilon_0 \) is the vacuum permittivity, and \( V \) represents the volume of the system. 

In this equation, \( \langle M^2 \rangle \) denotes the mean square of the total dipole moment, while \( \langle M \rangle^2 \) represents the square of the average dipole moment. This relationship highlights the connection between molecular fluctuations and the macroscopic dielectric response of the liquid.

\subsection{Finite system Kirkwood g-factor, G$_k$.}

Given that the orientations of molecules are crucial to dielectric properties, we have assessed the polarization factor Gk. This factor quantifies the equilibrium fluctuations of the system's collective dipole moment and is intrinsically linked to the orientational correlation function. Kirkwood theorised that it is possible to express the dielectric constant $\epsilon$ in terms of a short-range orientational correlation function \cite{kirk}. Accordingly, we evaluate the polarization factor Gk within this theoretical framework \cite{Glattli}.


 \begin{equation} 
 \label{Gk} 
 G_K = \frac{<\mathbf{M}^2>}{N \mu^2} \end{equation}
 

\begin{figure}[h!] \centering {\includegraphics[width=0.7\textwidth, angle=-90]{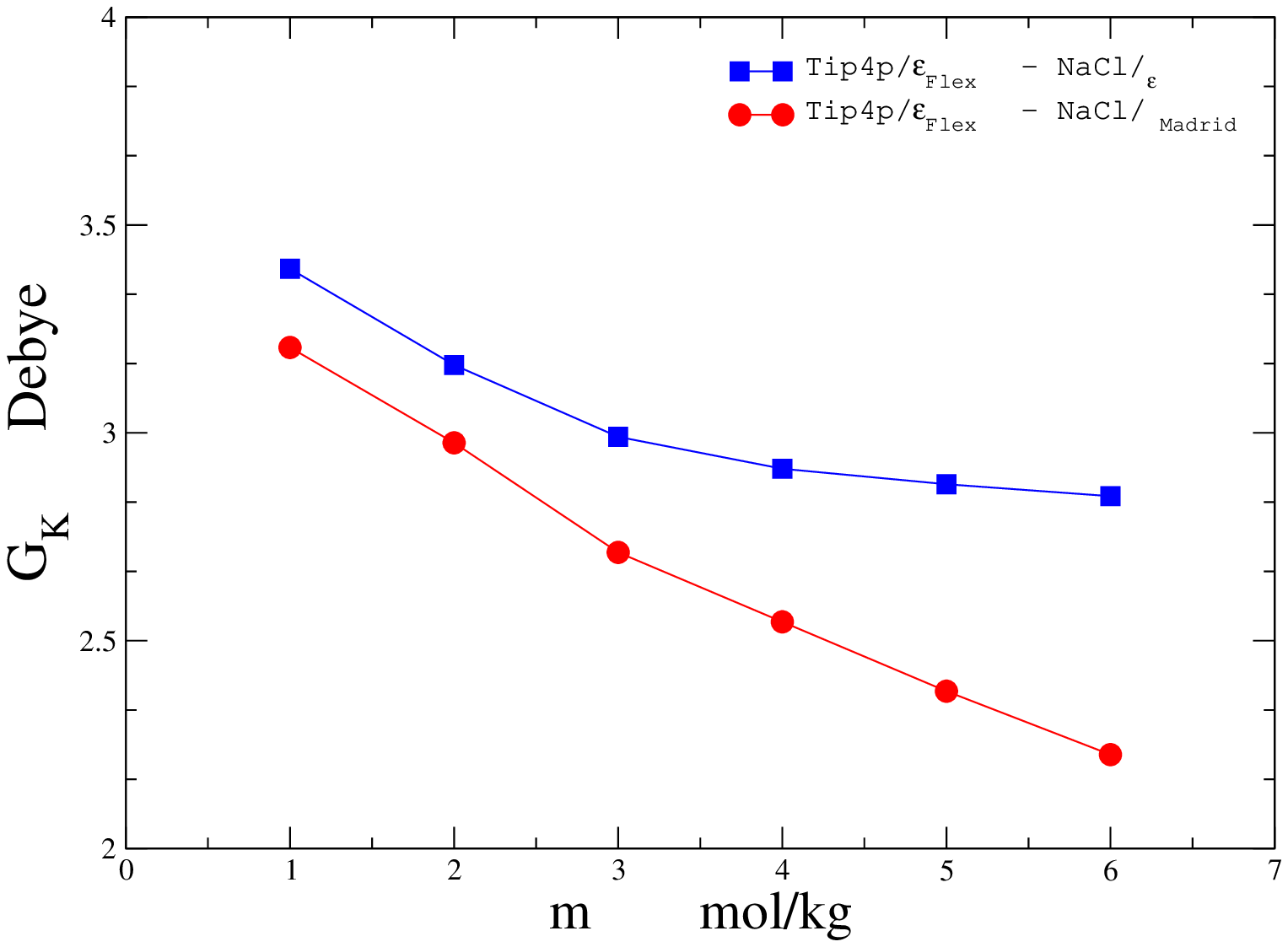}}
 \caption{The finite system Kirkwood g-factor, Gk, .}
 \label{Gk}

 \end{figure} 
 

Here, N denotes the total number of molecules, and M represents the total dipole moment $\mu$ in the system. It is noteworthy that local orientational correlations are averaged out by thermal motion after the first few coordination shells.

\subsection{Self Diffusion Coefficient,D}.

\begin{figure}[h]
	\centerline{\includegraphics[width=12.0cm,angle=-90]{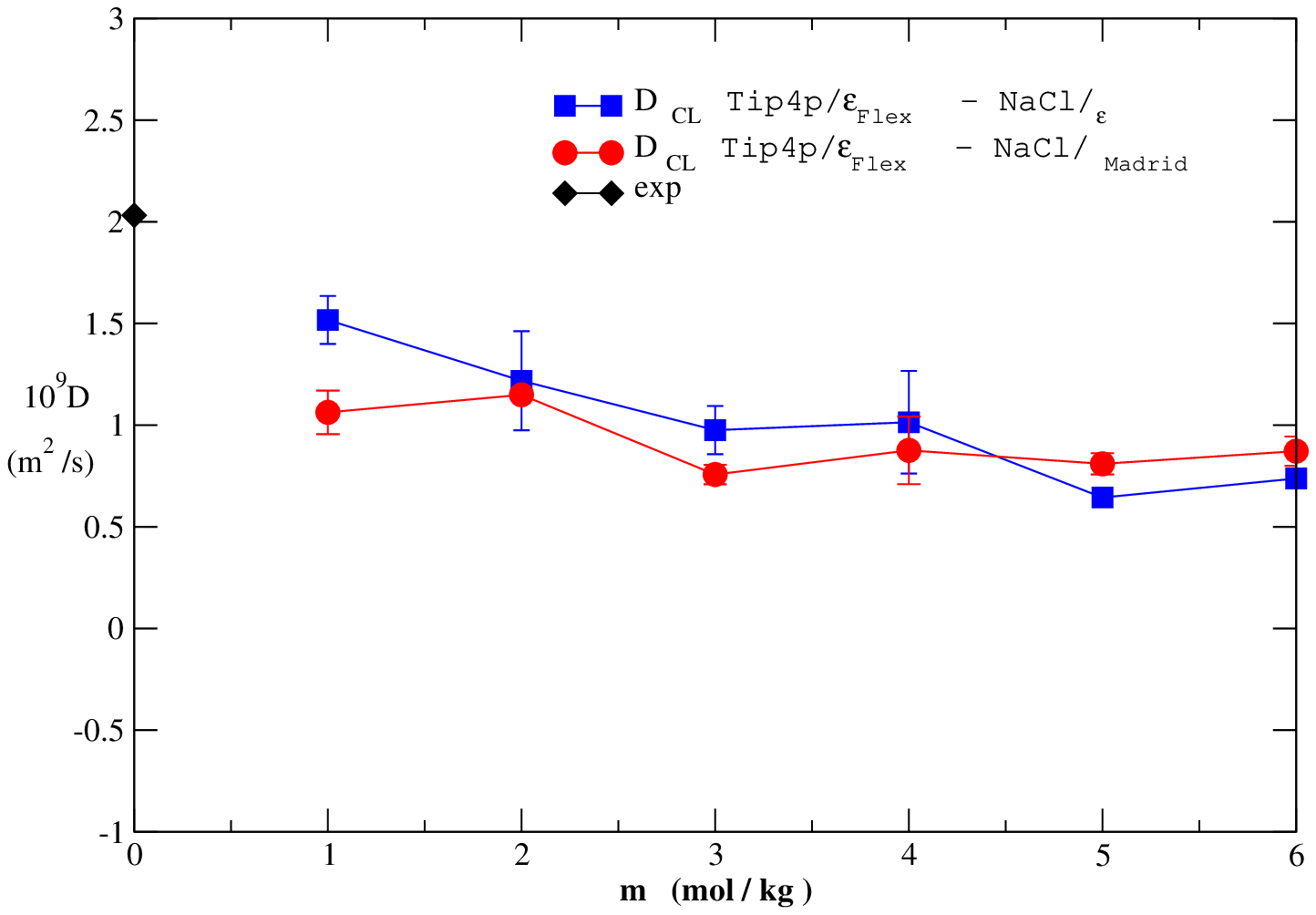}}
	\caption{Self Diffusion coefficient of chloride ions as a function of the molal concentration of NaCl at room temperature and pressure. The filled black diamond represents the experimental data~\cite{CRC}, , for NaCl/$\epsilon$ in TIP4P/$\epsilon_{Flex}$water (blue squares) and NaCl/Madrid in TIP4P/$\epsilon_{Flex}$, water (red circles)}
	\label{DCl}
\end{figure}

The self diffusion coefficient of chloride ions is depicted in Figure \ref{DCl}. The experimental data at infinite dilution indicates a diffusion coefficient of $D_{\text{Cl}} = 2.032 \times 10^{-5}$ cm$^2$ s$^{-1}$. We observed that, irrespective of the ionic model employed, the discrepancies in the results are minimal, and the calculated values are in close agreement. As anticipated, the system demonstrates a reduction in mobility as the concentration of ions increases. This behaviour is characteristic of molecular systems, where the increasing concentration leads to more frequent interactions and collisions between particles, ultimately impeding their movement.

The reduction in mobility is a direct consequence of the increased number of particles, which enhances intermolecular forces and collisions. This phenomenon is particularly pronounced in ionic solutions, where electrostatic interactions between charged particles and solvent molecules further hinder their diffusion. As a result, higher concentrations produce a more constrained dynamic environment, reducing the overall mobility of the ions within the solution.

\begin{figure}[h]
	\centerline{\includegraphics[width=12.0cm,angle=-90]{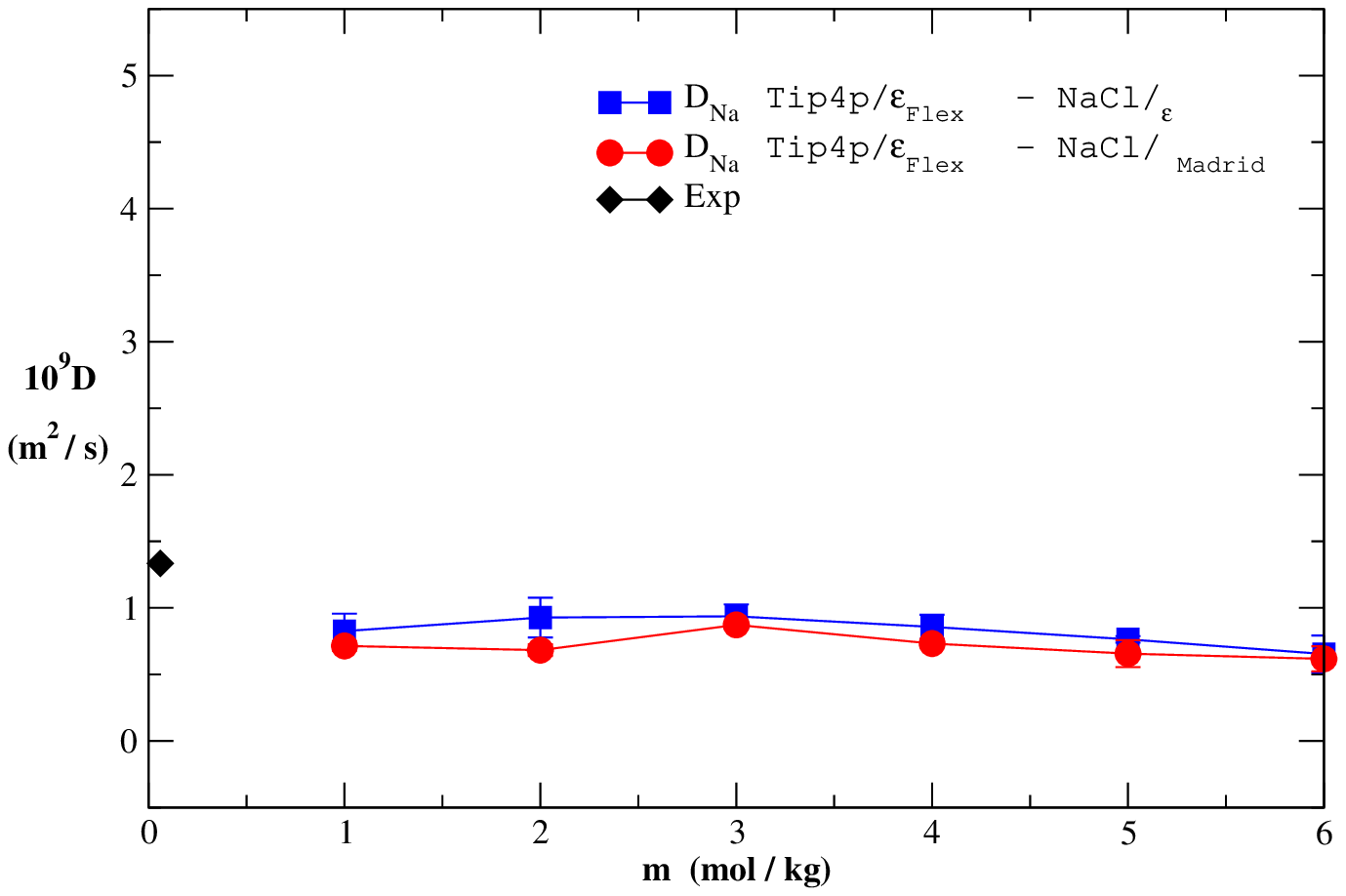}}
	\caption{Self Diffusion coefficient of sodium ions as a function of the molal concentration of NaCl at room temperature and pressure. The filled black diamond represents the experimental data~\cite{CRC}, , for NaCl/$\epsilon$ in TIP4P/$\epsilon_{Flex}$water (blue squares) and NaCl/Madrid in TIP4P/$\epsilon_{Flex}$, water (red circles)}
	\label{Dna}
\end{figure}

In Figure \ref{Dna}, the self diffusion coefficient of sodium ions is plotted against the molal concentration of NaCl. Unlike chloride ions, the diffusion coefficient of sodium remains relatively constant with increasing salt concentration. This observation suggests that the process is primarily governed by the Cl$^{-}$ ions, whose higher electronegativity significantly influences the dipole moment of the surrounding water molecules. This, in turn, reduces the dielectric constant of the solution and increases its density. The Na$^{+}$ ions contribute less to the dynamics of the mixture due to their lower electronegativity.

The experimental diffusion coefficient for sodium ions at infinite dilution is reported as $D_{\text{Na}} = 1.334 \times 10^{-5}$ cm$^2$ s$^{-1}$. 
\newpage

\subsection{Pressure and Temperature, P \& T}.

\begin{figure}[h!] \centering {\includegraphics[width=0.7\textwidth, angle=-90]{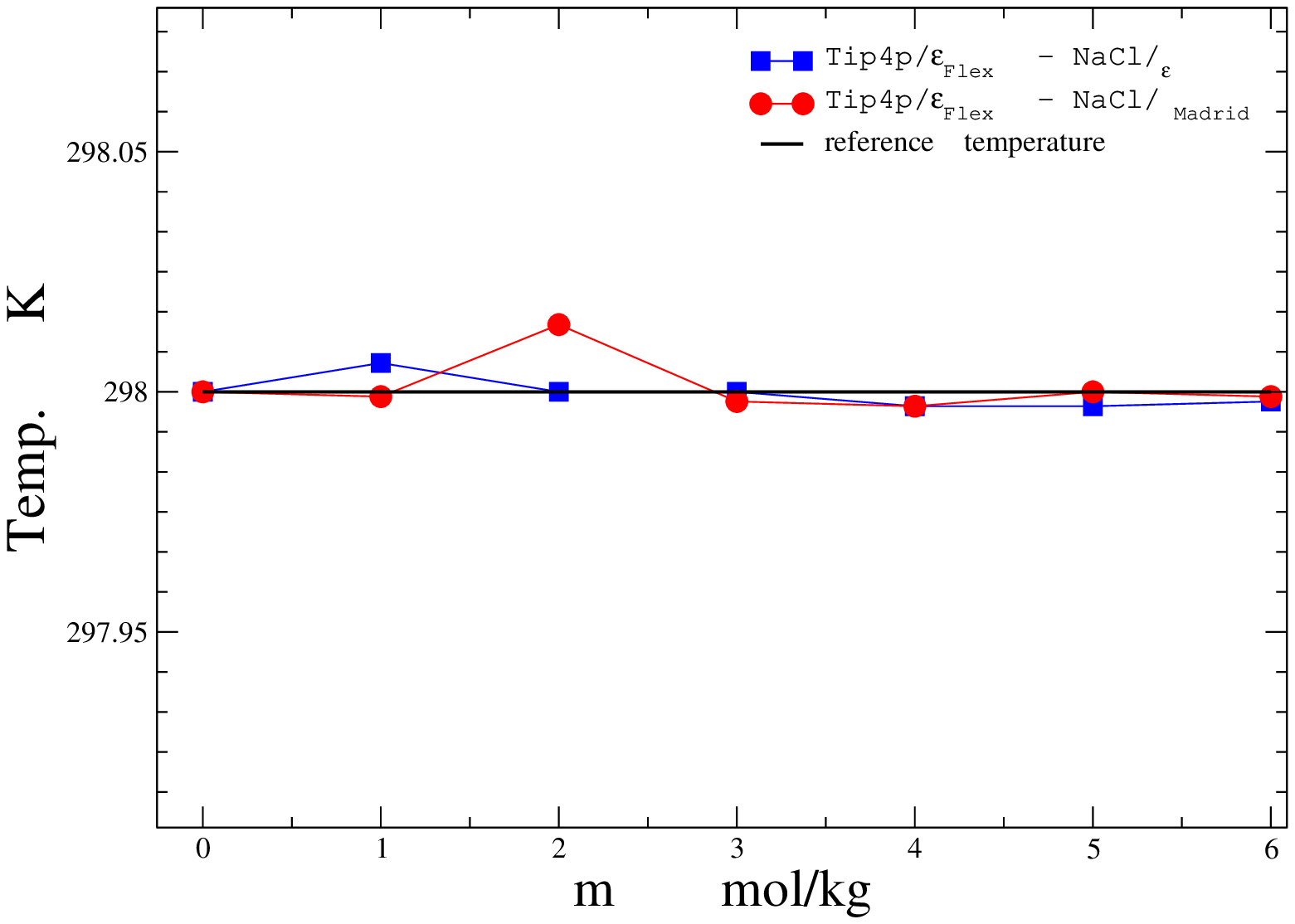}}
 \caption{}
 \label{temp}

 \end{figure}

As part of this study, we analysed the stability provided by the Nos\'e-Hoover thermostat. In Figure \ref{temp}, the temperature is plotted as a function of molal concentration after 150 ns of simulation.

The Nos\'e-Hoover thermostat was selected for its robust ability to maintain temperature in molecular dynamics simulations, effectively replicating the conditions of a canonical ensemble. The results presented in Figure \ref{temp} allow us to evaluate the efficiency of the thermostat in keeping the system stable at the target temperature throughout the simulation, even at varying salt concentrations. This aspect is crucial, as maintaining accurate temperature control ensures that the thermodynamic properties derived from the simulation are reliable and accurately reflect the behaviour of the system under the prescribed conditions.

To maintain the pressure, we employed the Parrinello-Rahman barostat, which stabilises the system by allowing volume fluctuations, ensuring an average pressure of 1 bar throughout the study.

\begin{figure}[h!] \centering {\includegraphics[width=0.7\textwidth, angle=-90]{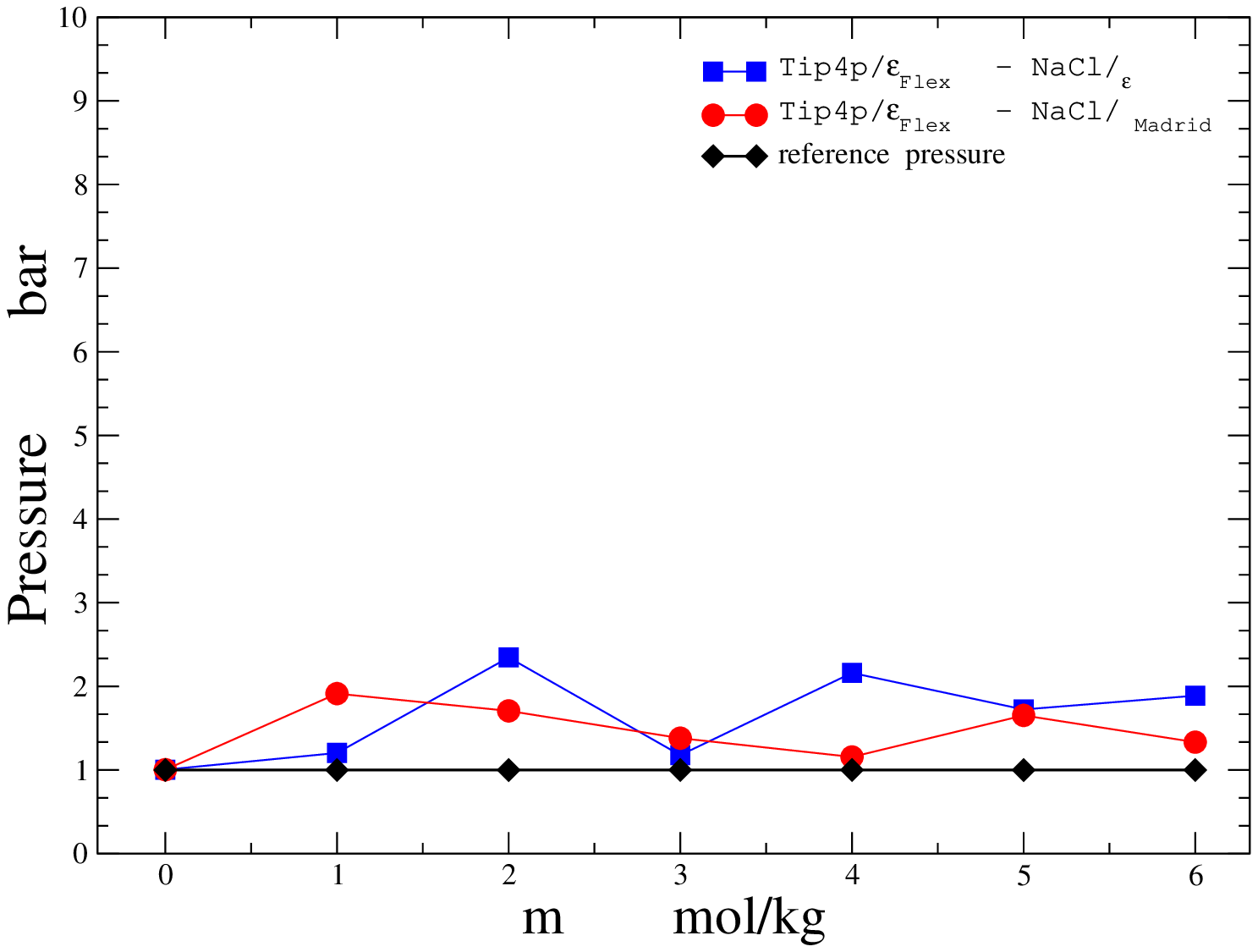}}
 \caption{}
 \label{pres}

 \end{figure}

As shown in Figure \ref{pres}, the pressure fluctuations in the case of NaCl/$\epsilon$ in TIP4P/$\epsilon_{Flex}$ are significantly larger, almost double, compared to the NaCl/Madrid in TIP4P/$\epsilon_{Flex}$. In the latter case, the fluctuations are smaller and tend to decrease as the molal concentration increases. This indicates that the system's pressure response varies depending on the force field employed, with the NaCl/$\epsilon$ model exhibiting a higher sensitivity to pressure changes, especially at higher concentrations.

\newpage

\section{Conclusions}

In this study, we investigated the dielectric and structural properties of water with varying concentrations of NaCl, using molecular dynamics (MD) simulations. We employed the TIP4P/$\epsilon_{Flex}$ water model alongside two distinct force fields for sodium chloride: NaCl/$\epsilon$ and NaCl/Madrid. The density is accurately reproduced by the NaCl/$\epsilon$ and TIP4P/$\epsilon_{Flex}$ models. However, in the case of the NaCl/Madrid and TIP4P/$\epsilon_{Flex}$ combination, there is a noticeable decrease in density at higher concentrations of [NaCl]. This reduction in density can be attributed to the underestimation of the dielectric constant, particularly in the NaCl/Madrid force field. Such discrepancies highlight the limitations of this model in capturing the correct dielectric properties at elevated salt concentrations, which, in turn, impacts its ability to simulate density accurately under these conditions.
\\
One key focus of the study was the dielectric properties of the electrolyte solution. We calculated the dielectric constant by examining fluctuations in the dipole moment, based on Neumann’s approach. Our results demonstrated that the NaCl/$\epsilon$ model provided a closer approximation to experimental data, especially at higher salt concentrations. In contrast, the NaCl/Madrid model tended to overestimate the dielectric constant, particularly at elevated temperatures. This discrepancy is largely attributed to weaker ion-water interactions in the NaCl/Madrid model, which results in less accurate representations of the system’s dielectric behaviour.

The clustering behaviour of ions was analysed, revealing that NaCl/$\epsilon$ formed larger clusters at higher salt concentrations, while NaCl/Madrid maintained smaller clusters with a higher proportion of free ions throughout the simulation. These cluster formations play a significant role in determining the solution’s structural properties.

Thermodynamic stability was another essential aspect of this work. The Nosé-Hoover thermostat effectively stabilised the system’s temperature over time, ensuring that the simulated thermodynamic properties closely resembled real-world conditions. In terms of pressure control, the Parrinello-Rahman barostat maintained stability across the simulations, though the NaCl/$\epsilon$ model exhibited larger pressure fluctuations compared to NaCl/Madrid. As molal concentration increased, these fluctuations tended to decrease, especially in the NaCl/Madrid model, where stability improved.

\newpage

\begin{acknowledgement}

RFA acknowledge support from CONAHCYT, Pollux computer at IF-BUAP and UAM  Supercomputing Center (Yotla) for the computer time allocated for some of our calculations.
	
\end{acknowledgement}

\bibliography{achemso}

\end{document}